\documentclass[12pt,a4paper]{article}

\title{\large \bf The Magueijo-Smolin Group is Linear in Five Dimensions }
\author{\normalsize Patrick J. McCarthy\footnote{e-mail: yhtraccm\_j\_p@hotmail.com} \\
\textit {\normalsize 123 Horn Lane, Woodford Green, Essex IG8 9AF, U.K.} 
 \\
\normalsize and \\
\normalsize R. L. Schafir\footnote{e-mail: r.schafir@herts.ac.uk}\\
\textit{\normalsize Department of Mathematics, University of Hertfordshire, Hatfield AL10 9AB, U.K.}}
\date{}
\begin{document}
\maketitle 
\begin{abstract}
\noindent Magueijo and Smolin have introduced a modification of the Lorentz group for the momentum-space
 transformations in Doubly-Special Relativity.  As presented the group is
non-linear, but we show that it is a group of fraction-linear transformations in 4 dimensional 
real projective space.  We pass to the associated 5 dimensional linear space and identify the subgroup 
as a conjugate of the ordinary Lorentz group, giving the 
conjugating matrix.  Taking the dual of the 5-space, we identify a subgroup
 as the Lorentz transformations about a point different to the origin.

\hfil\vskip\baselineskip \noindent PACS numbers: 03.30.+p, 04.50.+h,04.60.-m
\end{abstract}
\hfil\vskip\baselineskip
\hfil\vskip\baselineskip
\thispagestyle{empty}
\pagestyle{plain}
\par
In connection with Doubly-Special Relativity (for a review article see Ref [1]), Magueijo and Smolin [2] have introduced a group of transformations on momentum space.  These are:
\begin{equation}
\begin{array}{l} 
 p'_0  = {\displaystyle \frac{{\gamma (p_0  - vp_z )}}{{1 + l_P (\gamma  - 1)p_0  - l_P \gamma vp_z }}} \\ 
\\
 p'_x  = {\displaystyle \frac{{p_x }}{{1 + l_P (\gamma  - 1)p_0  - l_P \gamma vp_z }}} \\ 
\\
 p'_y  = {\displaystyle \frac{{p_y }}{{1 + l_P (\gamma  - 1)p_0  - l_P \gamma vp_z }}} \\ 
\\
 p'_z  = {\displaystyle \frac{{\gamma (p_z  - vp_0 )}}{{1 + l_P (\gamma  - 1)p_0  - l_P \gamma vp_z }}}\\ 

 \end{array}
\end{equation}
for a boost along the z-axis.  These equations are non-linear, but can be seen to be fractional-linear.  
First let us simplify the notation by writing $\lambda = l_p$ (the Plank length) and $v  = -\mathrm{tanh} \;\psi  $, where 
$\psi $ is the rapidity parameter.  We now write
\begin{equation}
p_\mu   = \frac{{\pi _\mu  }}{{\pi _4 }}
\end{equation}
where $\mu =0,1,2,3, \; p_1 = p_x, p_2 = p_y, p_3 = p_z$ and $\pi _4$ is an extra, fifth, coordinate.  A simple check then shows that the
transformation (1) is induced by the following linear transformation of the five $\pi $ variables.
\begin{equation}
\begin{array}{l}
  \pi '_0  = \pi_0 \cosh \psi  + \pi_3 \sinh \psi  
\\
 {\displaystyle \pi '_1  = \pi _1 }\\
{\displaystyle \pi '_2  = \pi _2 } \\ 
 {\displaystyle  \pi '_3  = \pi_3 \cosh \psi  + \pi _0 \sinh \psi} 
 
\\
 {\displaystyle \pi '_4  = \pi _4  + \lambda \left( {\pi _0 \cosh \psi  + \pi _3 \sinh \psi } \right) - \lambda \pi _0 } \\ 
 \end{array}
\end{equation}
which is described by the matrix
\begin{equation}
B(\psi ) = \left[ {\begin{array}{*{20}c}
   {\cosh \psi } & 0 & 0 & {\sinh \psi } & 0  \\
   0 & 1 & 0 & 0 & 0  \\
   0 & 0 & 1 & 0 & 0  \\
   {\sinh \psi } & 0 & 0 & {\cosh \psi } & 0  \\
   {\lambda \left( {\cosh \psi  - 1} \right)} & 0 & 0 & {\lambda \sinh \psi } & 1  \\
\end{array}} \right]
\end{equation}
\par
The transformations form a 1-parameter group.  If, similarly, we compute the matrices for x and y 
boosts and for spatial rotations, we can write the infinitesimal generator for the general element
 of the 6-dimensional Lie algebra:
\begin{equation}
l_\lambda = \left[ {\begin{array}{*{20}c}
   0 & a & b & c & 0  \\
   a & 0 & p & q & 0  \\
   b & { - p} & 0 & r & 0  \\
   c & { - q} & r & 0 & 0  \\
   0 & {a\lambda } & {b\lambda } & {c\lambda } & 0  \\
\end{array}} \right]
\end{equation}
where $a,b,c,p,q,r$ are the parameters for boosts and rotations respectively.  The limiting case $\lambda =0$ gives the usual Lorentz algebra
$L_0$ in the first four coordinates.
\par
From direct calculation we find that $l_\lambda g_\lambda =g_\lambda l_0$, i.e. $l_\lambda =g_\lambda l_0g_\lambda ^{-1}$,
where the conjugator $g_\lambda $ is given by
\begin{equation}
g_\lambda   = \left[ {\begin{array}{*{20}c}
   1 & 0 & 0 & 0 & 0  \\
   0 & 1 & 0 & 0 & 0  \\
   0 & 0 & 1 & 0 & 0  \\
   0 & 0 & 0 & 1 & 0  \\
   \lambda  & 0 & 0 & 0 & 1  \\
\end{array}} \right]
\end{equation}
By exponentiation, the Magueijo-Smolin group is conjugate to the ordinary Lorentz transformations (in the first
four variables), by the same conjugator $g_\lambda $.
\par
For a geometrical characterisation of these 5-dimensional transformations, 
let us look at the dual space to the space of $\pi $'s, with dual transformations defined in the 
following way.  Given a transformation $M$ on a space, define a dual transformation $\tilde M$ by
\begin{equation}
\left<\tilde M(\xi),M(x)\right> = \left<\xi, x\right>
\end{equation}
for a covector $\xi $ and vector $x$, where $\left< \;,\;\right>$ represents the contraction between them.  This has the 
homomorphism property that  $\tilde M_1\tilde M_2$ corresponds to $M_1M_2$. 
  To find 
matrix expressions for $\tilde M$ and $M$, suppose we have bases
 $\varepsilon _i$ and $e_i$ in the dual and original spaces, and define
\begin{equation}
\left<\varepsilon _i, e_j \right> = A_{ij} 
\end{equation}
Then for any covector $\sum\limits_i {\xi _i \varepsilon _i } $ and  vector $ \sum\limits_i {x_i } e_i $,
it follows from linearity that \\ $\left<
\sum\limits_i {\xi _i \varepsilon _i } \;, \; \sum\limits_j {x_j e_j }\right>  = \sum\limits_{ij} {\xi _i x_j A_{ij} }
 $ , $\;$
 i.e.
\begin{equation}
\left< \xi, x \right> = \xi ^{\rm{T}} Ax
\end{equation}
in matrix notation.
Then by (7) and (9) and the arbitrariness of $\xi $ and $x$ it follows that
\begin{equation}
\tilde M = (AM^{ - 1} A^{ - 1} )^{\rm{T}} 
\end{equation}
\par
In orthodox special relativity $A$ is the Minkowski $\eta = \rm{diag}(1,-1,-1,-1)$, so for the 5-dimensional 
case let us extend this minimally by choosing for $A$:
\begin{equation}
A = A^{ - 1}  = \left[ {\begin{array}{*{20}c}
   \eta  & {0^{\rm{T}} }  \\
   0 & 1  \\
\end{array}} \right]
\end{equation}
(writing the matrix in block diagonal form).
\par
The general form of the transformations on $\pi $-space may be written:
\begin{equation}
\left[ \begin{array}{l}
 \pi  \\ 
 \pi _4  \\ 
 \end{array} \right] \to \left[ {\begin{array}{*{20}c}
   \Lambda  & {0^{\rm{T}} }  \\
   {\lambda \varepsilon (\Lambda  - I)} & {\rm{1}}  \\
\end{array}} \right]\left[ \begin{array}{l}
 \pi  \\ 
 \pi _4  \\ 
 \end{array} \right]
\end{equation}
where $\Lambda $ is a Lorentz transformation on the first 4 coordinates, and ${\mathbf \varepsilon}$
is the row vector $(1,0,0,0)$.  (This is the generalisation of (4) and can be verified using exponentiation and 
conjugation with (5) and (6).)  If $M$ is the matrix in (12), it can be verified that
\begin{equation}
M^{ - 1}  = \left[ {\begin{array}{*{20}c}
   {\Lambda ^{ - 1} } & {0^{\rm{T}} }  \\
   {\lambda \varepsilon \left( {\Lambda ^{ - 1}  - I} \right)} & 1  \\
\end{array}} \right]
\end{equation}
Then by (10), and the facts that
\begin{equation}
\Lambda ^T\eta =\eta \Lambda ^{-1}\:,\;\;\eta \Lambda ^{-1}\eta =\Lambda ^T\:,\;\;\eta \Lambda ^T=\Lambda ^{-1}\eta 
\end{equation}
and
\begin{equation}
\lambda \varepsilon (\Lambda ^{ - 1}  - I)\eta  = \lambda \eta \varepsilon (\Lambda ^{\rm{T}}  - I) = \lambda \varepsilon (\Lambda ^{\rm{T}}  - I)
\end{equation}
we obtain
\begin{equation}
\tilde M =  \left[ {\begin{array}{*{20}c}
   \Lambda  & {\lambda (\Lambda  - I)\varepsilon ^{\rm{T}} }  \\
   0 & 1  \\
\end{array}} \right]
\end{equation}
\par
Thus the transformations dual to (12) may be written
\begin{equation}
\left[ \begin{array}{l}
 x \\ 
 x^4  \\ 
 \end{array} \right] \to \left[ {\begin{array}{*{20}c}
   \Lambda  & {\lambda (\Lambda  - I)\varepsilon ^{\rm{T}} }  \\
   0 & {\rm{1}}  \\
\end{array}} \right]\left[ \begin{array}{l}
 x \\ 
 x^4  \\ 
 \end{array} \right]
\end{equation}
These leave $x^4$ fixed, while if $x^4 = 1$ the other coordinates transform by
\begin{equation}
x \to \Lambda (x + \lambda \varepsilon ^{\rm{T}} ) - \lambda \varepsilon ^{\rm{T}} 
\end{equation}
This is a Lorentz transformation, though not about the origin but about the ``Plank-time" point
 $-\lambda \varepsilon^T$.  Lorentz transformations about this point, and the Lorentz
transformations about the origin, are  subgroups of the Poincar$ \acute e$ group conjugate to each other
by a translation.
\par
Since momentum space is the cotangent space (with cotangents at different points identified
 using the trivial connection on the flat spacetime), the dual is the tangent space, and we
 see that the above subgroup when acting in four dimensions is a non-linear (because not origin 
preserving) set of transformations on the tangent space at an arbitarily chosen origin in the 
underlying position space.  These cannot be the induced maps to maps on the position space which
 leave the chosen point fixed (since the induced map is linear), and this illustrates the 
difficulty in finding suitable 
position space transformations  in four dimensions, even non-linear ones.
\par
The fact that $x^4$ is left invariant by the Lorentz-type subgroup suggests that the fifth coordinate 
may be connected with the proper mass, something which appears in orthodox special relativity only
 as an external parameter.  If it does have this significance then mass is entering the formalism 
just as time previously entered the formalism, changing its status from external parameter to an
 extra degree of freedom.  
More general transformations in which the fifth coordinate is changed 
would then correspond to a freedom to change the proper mass.

\end{document}